\documentclass[11pt]{article}
\usepackage{graphicx,color,bm}
\usepackage{fancyhdr,lastpage}
\usepackage{subeqn}
\usepackage{slashed}
\def\simgt{\,{\rlap{\lower 3.5pt\hbox{$\mathchar\sim$}}\raise 1pt\hbox{$>$}}\,}
\def\simlt{\,{\rlap{\lower 3.5pt\hbox{$\mathchar\sim$}}\raise 1pt\hbox{$<$}}\,}
\def\fds{f_{D_s}}
\def\ds{D_s}

\begin{document}
\thispagestyle{empty}
\vspace*{-15mm}
%----------
\baselineskip 10pt
\begin{flushright}
\begin{tabular}{l}
{\bf OCHA-PP-303}\\
\end{tabular}
\end{flushright}
\baselineskip 24pt 
\vglue 10mm 
%%%%%%%%%%%%%%%%%%%%%%%%%%%%%%%%%%%%%%%%%%%%%%
% Title
%%%%%%%%%%%%%%%%%%%%%%%%%%%%%%%%%%%%%%%%%%%%%%
\begin{center}
{\Large\bf
Leptonic decays of $D_s$ and $B^+$ mesons in 
supersymmetric standard model with $R$-parity violating 
interactions
}
\vspace{5mm}

\baselineskip 18pt
\def\thefootnote{\fnsymbol{footnote}}
\setcounter{footnote}{0}
{\bf
Yumiko Aida$^{1)}$, Eri Asakawa$^{2)}$, Gi-Chol Cho$^{2)}$, 
Hikaru Matsuo$^{1)}$
\vspace{5mm}

$^{1)}${\it 
Graduate School of Humanities and Sciences,
Ochanomizu University, Tokyo, 112-8610, Japan
}\\
$^{2)}${\it 
Department of Physics, Ochanomizu University, Tokyo 112-8610, Japan}\\
}
\vspace{10mm}
\end{center}

%%%%%%%%%%%%%%%%%%%%%%%%%%%%%%%
%%%%%                     %%%%%
%%%%%      Abstract       %%%%%
%%%%%                     %%%%%
%%%%%%%%%%%%%%%%%%%%%%%%%%%%%%%
\begin{center}
{\bf Abstract}\\[7mm]
\begin{minipage}{14cm}
\baselineskip 18pt
\noindent
%%%%%----------------------------------
We investigate leptonic decays $D_s \to \tau \nu_\tau$ and 
$B^+ \to \tau \nu_\tau$ in $R$-parity violating (RPV) supersymmetric  
standard model. 
Taking account of interference between the $s$-channel 
slepton exchange and the $t$-channel squark exchange diagrams, 
we find that the supersymmetric contributions are cancelled between 
two diagrams so that the RPV couplings could be sizable under 
the experimental bounds. 
Constraints on the relative sign between the RPV couplings in $s$- 
and $t$-channel diagrams are also discussed. 
%%%%%----------------------------------
\end{minipage}
\end{center}
\newpage

\baselineskip 18pt
\section{Introduction}
Supersymmetric extension of the standard model 
(SM)~\cite{Martin:1997ns} is a leading candidate of physics beyond 
the SM.  
However, since no experimental evidence of supersymmetry (SUSY) has 
not been found yet, discovery of supersymmetric particles at energy 
frontier experiments such as LHC is one of the important tasks of 
particle physics. 
%%%

%%%
The most general gauge invariant and renormalizable superpotential 
in the supersymmetric SM 
contains baryon ($B$) and lepton ($L$) number violating interactions
which may lead to unwanted fast proton decay or sizable lepton number
violating processes. 
Such interactions can be forbidden by introducing so called $R$-parity 
which is defined as $R=(-1)^{3B+L+2S}$, where $S$ denotes the spin
quantum number.  
Owing to the $R$-parity, in addition to the suppression of 
$B$- and $L$-violating processes, the lightest supersymmetric particle
(LSP) becomes stable, and it could be a candidate of dark matter. 
On the other hand, some of $R$-parity violating (RPV) interactions may  
play phenomenologically attractive role. 
For example, the $R$-parity and $L$-violating interactions may explain 
tiny neutrino mass without introducing the right-handed 
neutrinos~\cite{Hall:1983id}. 
Also, a possibility of gravitino dark matter due to the $R$-parity
violating interactions has been discussed 
in ref.~\cite{Buchmuller:2007ui}.  
%%% 

%%%
In this article, we study contribution of the RPV interactions 
to the leptonic decays of $D_s$ and $B^+$ mesons.  
It is known that the experimental data of the leptonic decays  
of $D_s$ and $B^+$ mesons slightly deviate from the SM expectations. 
Comparison of the experimental results of leptonic decay of $D_s$ meson  
is often presented in terms of the decay constant $f_{D_s}$.  
The recent measurement of the leptonic decay of $\ds$ meson  
by CLEO~\cite{Naik:2009tk} is given by
\begin{eqnarray}
\fds = 259.0 \pm 6.2 \pm 3.0 = 
259.0 \pm 6.9~{\rm [MeV]}, 
\label{cleo2009}
\end{eqnarray}
while the most precise calculation of $\fds$ by HPQCD and 
UKQCD~\cite{Follana:2007uv} is given as 
\begin{eqnarray}
\fds = 241 \pm 3 {\rm [MeV]}. 
\label{fdqcd}
\end{eqnarray}
The discrepancy between (\ref{cleo2009}) and (\ref{fdqcd}) is 
about $2.4\sigma$. % ($7\%$). 
The recent review of the experimental data and theoretical estimations 
on the decay constant $\fds$ can be found 
in refs.~\cite{Kronfeld:2009cf}.  
%%%

%%%
For the leptonic decay $B^+ \to \tau \nu_\tau$ , 
the experimental data of the branching ratio have been given 
by Belle and BABAR~\cite{ikado,btau}. 
The average of the data given by the UTfit
collaboration~\cite{Bona:2009cj} is 
\begin{eqnarray}
{\rm BR}(B^+ \to \tau \nu_\tau)_{\rm exp} &=& 
(1.73 \pm 0.34)\times 10^{-4}, 
\label{expB}
\end{eqnarray}
while the SM prediction is given by~\cite{Bona:2009cj}
\begin{eqnarray}
{\rm BR}(B^+ \to \tau \nu_\tau)_{\rm SM} 
&=& (0.84 \pm 0.11) \times 10^{-4}. 
\label{utfit}
\end{eqnarray}
The difference between (\ref{expB}) and (\ref{utfit})
is $2.5\sigma$. 
The deviations in the leptonic decays in both $D_s$ and $B^+$ may be 
statistical fluctuations. 
However, another interpretation of the deviations is that the 
deviations are caused by new physics beyond the SM. 
In the SM, these leptonic decays are dominated by the $W$-boson exchange  
at tree level\footnote{It has been pointed out that the radiative
corrections are highly suppressed~\cite{Burdman:1994ip}}. 
Therefore, a class of new physics models which lead to the leptonic 
decays at tree level could be candidates to explain the discrepancies,  
e.g., Two Higgs doublet model~\cite{Ahn:2010zza, Akeroyd:2009tn, 
Akeroyd:2007eh}, 
leptoquark model~\cite{Dobrescu:2008er, Benbrik:2008ik}, 
and the $R$-parity violating supersymmetric 
SM~\cite{Baek:1999ch,Dreiner:2001kc,Dreiner:2006gu, 
Kundu:2008ui, Kao:2009mz, Bhattacharyya:2009hb}. 
%%% 

%%%
In the supersymmetric SM with RPV interactions, contributions 
to the leptonic decays of $D_s$ and  
$B^+$ mesons are given by down-squark exchange in $t$-channel diagram,  
charged slepton exchange in $s$-channel diagram and 
charged Higgs boson exchange in $s$-channel diagram. 
In refs.~\cite{Dreiner:2006gu,Kundu:2008ui, Bhattacharyya:2009hb}, 
only the $t$-channel contribution was examined based on some scenarios 
or single coupling dominance hypothesis. 
The contribution of $s$-channel diagram in addition to the $t$-channel 
has been studied in refs.~\cite{Baek:1999ch,Dreiner:2001kc,Kao:2009mz}. 
However, since works in refs.~\cite{Baek:1999ch,Dreiner:2001kc} 
have been done before the first measurement of $B^+ \to \tau \nu_\tau$ 
in 2006~\cite{ikado}, bounds on the RPV couplings were not obtained 
from the experimental data of the $B^+$ decay. 
In ref.~\cite{Kao:2009mz}, constraints on the RPV couplings 
in $s$- and $t$-channel diagrams were investigated separately, 
and no interference effect between two diagrams was examined. 
In our study, we investigate the supersymmetric contributions to the 
leptonic decay of $D_s$ and $B^+$ mesons taking account of 
the interference effects between the $s$- and $t$-channel diagrams. 
We also examine a diagram mediated by the charged Higgs boson in 
the $s$-channel. 
Taking account of the interference effects between the $s$- and
$t$-channel diagrams, we show allowed region of the 
RPV couplings which explains the deviation between 
experimental data and the SM prediction. 
Note that the interference between two diagrams could be either 
constructive or destructive due to the relative sign of the RPV 
couplings in two diagrams. 
We find, therefore, that the experimental data constrains not only 
the size of RPV couplings but also the relative sign between 
the RPV couplings in $s$- and $t$-channel diagrams, which has not been 
examined in previous studies. 
The contribution of charged Higgs boson is found to be negligible 
in $D_s \to \tau \nu_\tau$, but sizable in $B^+ \to \tau \nu_\tau$. 
We discuss how the constraints on RPV couplings are affected by 
the charged Higgs contribution. 
%%%
\section{Set up}
The $R$-parity violating interactions with trilinear couplings are
described by the following superpotential 
\begin{eqnarray}
W_{\slashed R} &=& 
\frac{1}{2}\lambda_{ijk} L_i L_j E_k 
+ \lambda'_{ijk} L_i Q_j D_k 
+ \frac{1}{2} \lambda''_{ijk} U_i D_j D_k,  
%+ \mu' L H_u
\label{rparity}
\end{eqnarray}
where $Q$ and $L$ are SU(2)$_L$ doublet quark and lepton superfields, 
respectively. 
The up- and down-type singlet quark superfields are represented by 
$U$ and $D$, while the lepton singlet superfield is $E$. 
The generation indices are labeled by $i,j$ and $k$. 
The SU(2)$_L$ and SU(3)$_C$ gauge indices are suppressed. 
The coefficient $\lambda_{ijk}$ is anti-symmetric for $i$ and $j$, 
while $\lambda''_{ijk}$ is anti-symmetric for $j$ and $k$. 
For a comprehensive review of the $R$-parity violating supersymmetric
SM, see, ref.~\cite{Barbier:2004ez}. 
Constraints on the RPV couplings 
$\lambda_{ijk}, \lambda'_{ijk}$ and $\lambda''_{ijk}$ from various 
processes have been studied in the literature~
\cite{Barger:1989rk, Dreiner:1997uz, Bhattacharyya:1997vv, 
Allanach:1999ic, Dreiner:2006gu}. 
Since the baryon number violating coupling $\lambda''_{ijk}$ 
induces too fast proton decay, we take $\lambda''_{ijk}=0$ in 
the following.  
Then, the leptonic decays of $D_s$ and $B^+$ mesons occur through the 
$t$-channel exchange with a product of two $\lambda'$ couplings while 
$s$-channel exchange is given by a product of $\lambda$ and $\lambda'$. 
%
% new paragraph

%
Let us briefly summarize the leptonic decay of a pseudo scalar meson $P$ 
which consists of the up and (anti-) down-type quarks $u_a$ and  
$\bar{d_b}$, where $a,b$ are generation indices of quarks. 
The decay width of $P \to l_i \nu_j$ is given as 
\begin{eqnarray}
\Gamma(P \to l_i \nu_j ) 
&=& \frac{1}{8\pi} r_P^2 G_F^2 |V_{u_a d_b}^*|^2 f_P^2 m_{l_i}^2
m_P \left(1-\frac{m_{l_i}^2}{m_P^2}\right)^2
\label{dw}
\end{eqnarray}
where $G_F,V_{u_a d_b},m_{l_i}$ and $m_P$ are the Fermi constant, the 
Cabibbo-Kobayashi-Maskawa matrix element, the mass of a charged lepton 
$l_i$ and the mass of a pseudo scalar meson $P$, 
respectively. 
The flavor indices of charged leptons and neutrinos are expressed 
by $i$ and $j$, respectively. 
The decay constant is denoted by $f_P$. 
A parameter $r_P$ is defined as, 
\begin{eqnarray}
r_P^2 &\equiv& \frac{\left|G_F V_{u_a d_b}^* + A^P_{ii} \right|^2}
{ G_F^2 \left|V_{u_a d_b}^*\right|^2}
+
\sum_{j(\neq i)}
\frac{\left|A^P_{ij} \right|^2}
{ G_F^2 \left|V_{u_a d_b}^*\right|^2}, 
\label{rparam}
\end{eqnarray}
where $A^P_{ij}$ represents new physics contribution. 
Note that, in the second term of r.h.s. in (\ref{rparam}), 
one should take a sum only for $j$ (neutrinos), because that 
the neutrino flavor cannot be detected experimentally. 
If there is no new physics contribution, $r_P=1$. 
The interaction Lagrangian of the $t$-channel contribution 
to the decay width (\ref{dw}) can be obtained from 
the superpotential (\ref{rparity});  
\begin{eqnarray}
{\cal L} &=& 
\lambda'_{ijk} \left\{
-\overline{(l^c_L)_i} (u_L)_j (\widetilde{d}_R)^*_k
\right\}
+ 
\lambda'^*_{ijk} \left\{
(\widetilde{d}_R)_k
\overline{(d_L)_j} 
(\nu^c_L)_i
\right\} + {\rm h.c.}. 
\label{tch}
\end{eqnarray} 
Using the Fierz transformation, 
the effective Lagrangian which describes the $t$-channel  
squark exchange is given as 
\begin{eqnarray}
{\cal L_{\rm eff}^{\it t}} &=& 
\frac{1}{8}\sum_{k=1}^3
\frac{\lambda'_{iak} \lambda'^*_{jbk}}
{m_{\widetilde{d}_{Rk}}^2} ~
\bar{\nu}_j \gamma^\mu (1-\gamma_5) l_i~
\bar{d}_b \gamma_\mu (1-\gamma_5) u_a. 
\end{eqnarray}
For comparison, we show the effective Lagrangian for the $W$-boson 
exchange 
\begin{eqnarray}
{\cal L_{\rm eff}^{\rm SM}} &=& 
\frac{G_F}{\sqrt{2}} V_{u_a d_b}^* 
\bar{\nu}_i \gamma^\mu (1-\gamma_5) l_i~
\bar{d}_b \gamma_\mu (1-\gamma_5) u_a. 
\end{eqnarray}
Using the decay constant $f_P$ which is given by 
\begin{eqnarray}
\langle 0 | \bar{d}_b \gamma^\mu \gamma_5 u_a| P(q) \rangle 
= i f_P q^\mu, 
\label{decayconst}
\end{eqnarray}
we find the $t$-channel squark contribution to 
the decay $P(u_a \bar{d}_b) \to l_i \nu_j$ as  
\begin{eqnarray}
(A_t^P)_{ij} &=& \frac{1}{4\sqrt{2}}
\sum_{k=1}^3
\frac{\lambda'_{iak} \lambda'^*_{jbk}}
{m_{\widetilde{d}_{Rk}}^2}. 
\label{atp}
\end{eqnarray}
The $s$-channel contribution can be calculated from the interaction 
Lagrangian 
\begin{eqnarray}
{\cal L} &=& \lambda_{ijk} 
\left\{-\overline{(l_R)_k} (\nu_L)_j (\widetilde{l}_L)_i
\right\}
+ 
\lambda'_{ijk}\left\{
-\overline{(d_R)_k}(u_L)_j (\widetilde{l}_L)_i
\right\}
+ {\rm h.c.}. 
\end{eqnarray}
The effective Lagrangian is given by 
\begin{eqnarray}
{\cal L^{\it s}_{\rm eff}} &=& -\frac{1}{4}
\sum_{k=1}^3
\frac{\lambda^*_{kji}\lambda'_{kab}}{m_{\widetilde{l}_{Lk}}^2}
\bar{\nu}_j (1+\gamma_5) l_i ~
\bar{d}_b (1-\gamma_5) u_a. 
\end{eqnarray}
From (\ref{decayconst}) and equations of motion for $u,d$ quarks, 
we find 
\begin{eqnarray}
\langle 0| 
\bar{d}_b \gamma_5 u_a | P(q) \rangle 
= -i \frac{m_P^2}{m_{u_a}+m_{d_b}} f_P. 
\label{psdecay}
\end{eqnarray}
Using (\ref{psdecay}), we obtain the $s$-channel contribution as 
\begin{eqnarray}
(A_s^P)_{ij} &=& -\frac{1}{2\sqrt{2} m_{l_i}} 
 \frac{m_P^2}{m_{u_a}+m_{d_b}} 
\sum_{k=1}^3 
\frac{\lambda^*_{kji}\lambda'_{kab}}{m_{\widetilde{l}_{Lk}}^2}. 
\label{asp}
\end{eqnarray}
The charged Higgs contribution can be calculated from the 
interaction Lagrangian, 
\begin{eqnarray}
{\cal L} &=& V_{u_a d_b}^* 
\left\{
\frac{g m_{d_b}}{\sqrt{2}m_W} \tan\beta \overline{d_b} P_L u_a H^- 
+
\frac{g m_{u_a}}{\sqrt{2}m_W} \cot\beta \overline{d_b} P_R u_a H^- 
\right\}
\nonumber \\
&&~~~
+ 
\frac{g m_{l_i}}{\sqrt{2}m_W} \tan\beta \overline{\nu_i} P_R l_i H^+ 
+ {\rm h.c.}, 
\label{chhiggs}
\end{eqnarray}
where $g$ denotes the SU(2)$_L$ gauge coupling constant, and 
$\tan\beta \equiv \langle H_u \rangle/
\langle H_d \rangle$ is a ratio of the vacuum expectation values 
of two Higgs doublets $H_u$ (the weak hypercharge $Y=1/2$) 
and $H_d$ ($Y=-1/2$).  
We obtain the charged Higgs contribution $A_s^P$ from (\ref{chhiggs}) 
as 
\begin{eqnarray}
A^P_H &=& -G_F V_{u_a d_b}^* \frac{m_{d_b}}{m_{u_a} + m_{d_b}} 
\frac{m_P^2}{m_{H^-}^2}
\left(\tan^2\beta - \frac{m_{u_a}}{m_{d_b}}\right).  
\label{a_h}
\end{eqnarray}
Note that since leptons in the final state due to the charged Higgs 
exchange are flavor diagonal, the indices $i,j$ are suppressed in 
l.h.s. of (\ref{a_h}). 
%%%

%%%
%--------
\begin{figure}
 \begin{center}
  \includegraphics[width=7cm]{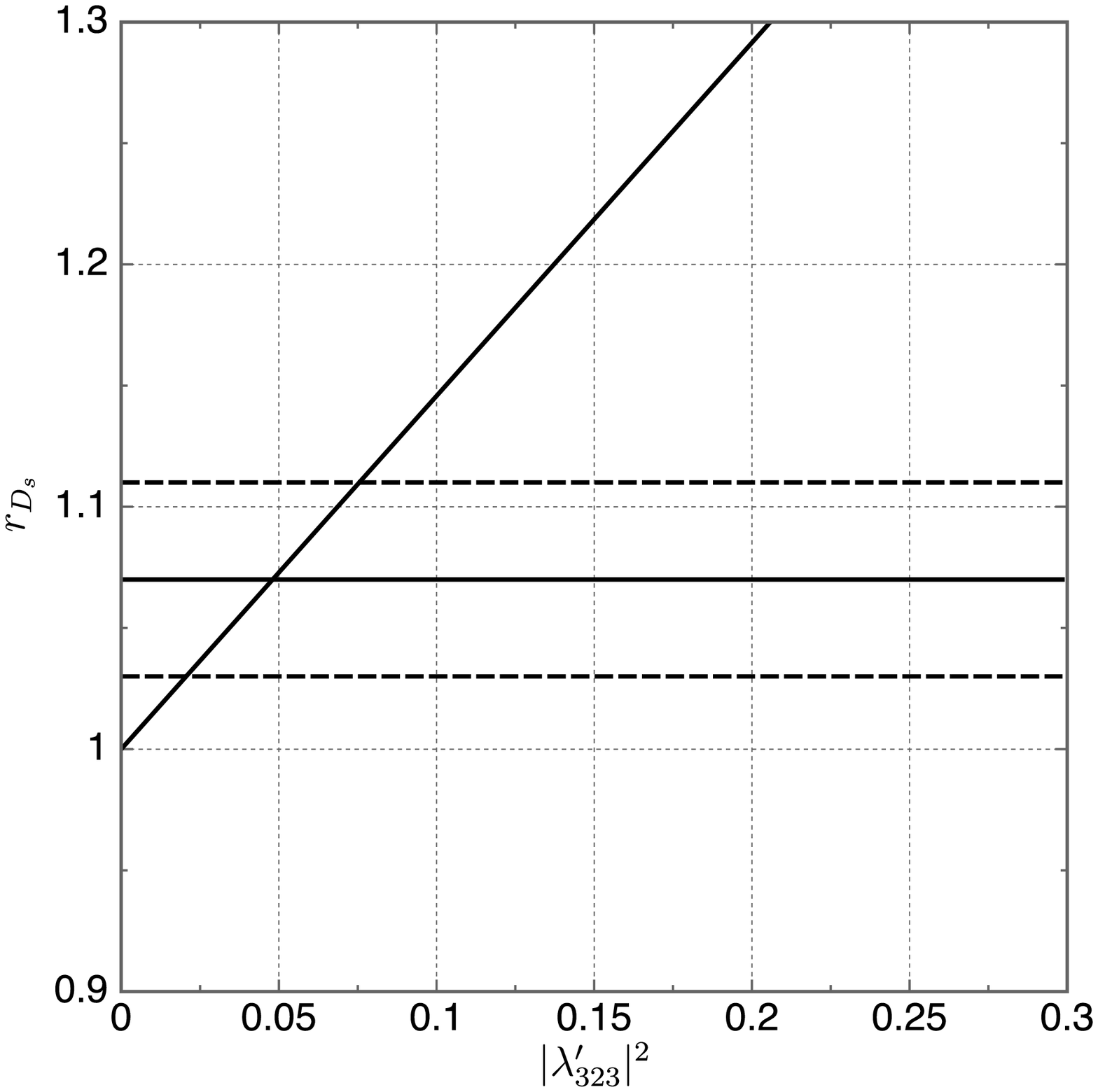}
  \includegraphics[width=7cm]{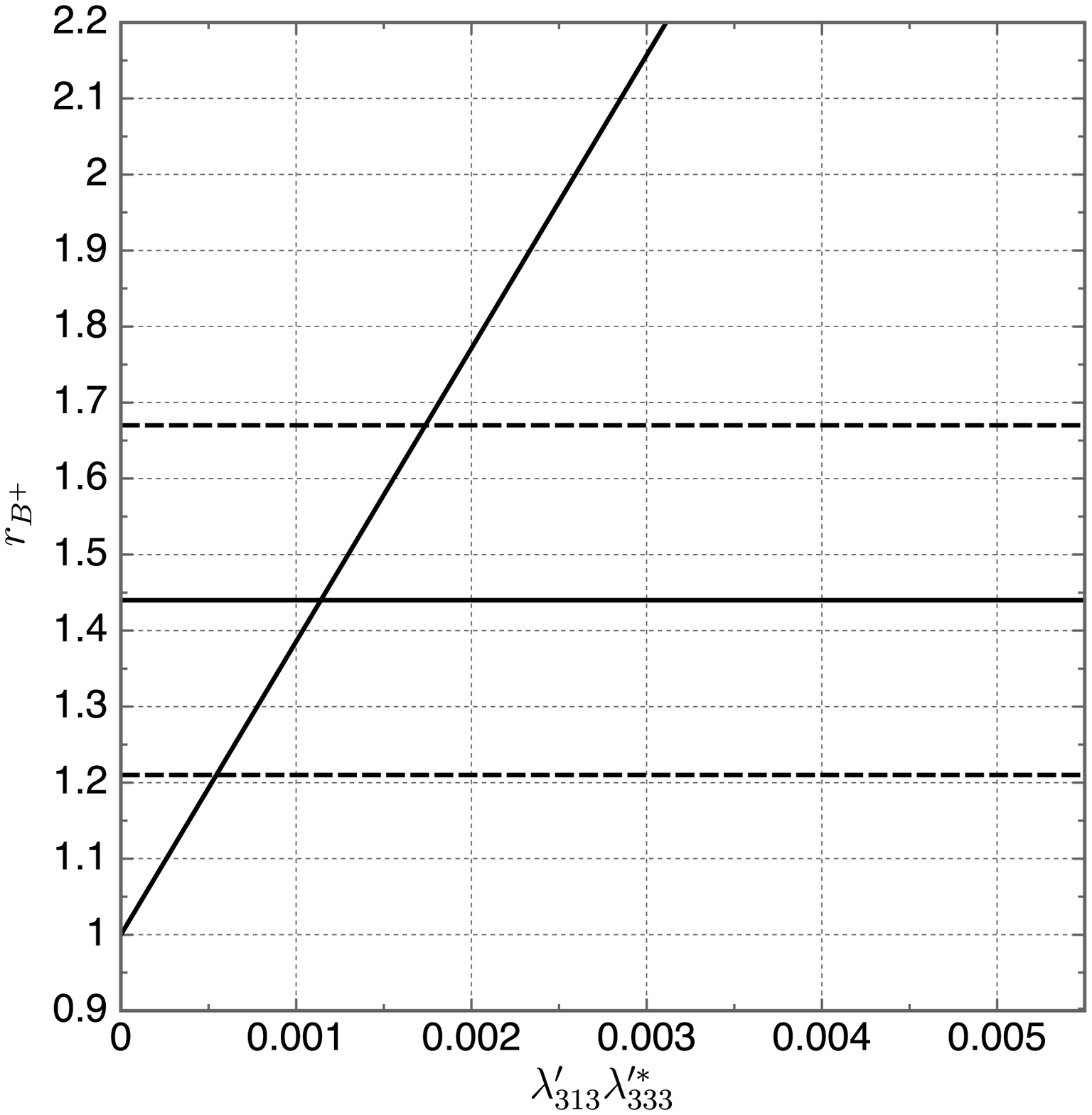}
  \caption{
Contribution of $t$-channel squark exchange 
to the $r$-parameters (\ref{rparam}) for $D_s \to \tau \nu_\tau$ (left)   
and $B^+ \to \tau \nu_\tau$ (right) as functions of the RPV couplings. 
The squark mass is fixed at $100~{\rm GeV}$. The horizontal lines 
denote the 1-$\sigma$ constraints on $r_{D_s}$ and $r_{B^+}$ 
given in eqs.~(\ref{rparamexp_ds}) and (\ref{rparamexp_bp}), 
respectively. 
\label{tch_graph}
}
 \end{center}
\end{figure}

\section{Numerical Study}
Next we examine the RPV contributions to the leptonic decays 
$P\to \tau \nu_\tau$ ($P=D_s$ or $B^+$) numerically. 
In the numerical study, 
we adopt the central values of the following parameters~\cite{PDG2010} 
\begin{eqnarray}
|V_{cs}|&=&1.023\pm 0.036, 
~~|V_{ub}|=(3.89\pm 0.44)\times 10^{-3}, 
\nonumber \\
m_{D_s} &=& 1968.47 \pm 0.33~{\rm MeV}, ~~
m_{B^+} = 5279.17 \pm 0.29~{\rm MeV}, 
\end{eqnarray}
In the analysis, we drop the second term in r.h.s. of (\ref{rparam}), 
i.e., the flavor off-diagonal final state 
such as $\tau \nu_\mu$ or $\tau \nu_e$ are neglected. 
Since the RPV couplings responsible for $P \to \tau \nu_\mu$ or 
$P \to \tau \nu_e$ induce the lepton flavor violating processes  
$\tau \to \mu \gamma$  or $\tau \to e \gamma$, 
those couplings must be highly suppressed. 
Therefore we neglect the $\tau \nu_\mu$ and $\tau \nu_e$ channels 
in the following study, i.e., $A_{ij}^P=0$ for $i\neq j$. 
Throughout out study, the squark and slepton masses are fixed at
$100~{\rm GeV}$. 
For simplicity, 
in the $t$-channel diagram we consider the sbottom exchange. 
On the other hand, the stau exchange is forbidden in the $s$-channel
diagram, and we consider the smuon exchange. 
Let us recall that $s$-channel amplitude is proportional to 
a product of $\lambda$ and $\lambda'$. 
Since the final state is $\tau \nu_\tau$, the RPV coupling 
$\lambda_{i33}$ requires $i\neq 3$ due to the anti-symmetric property  
of $\lambda_{ijk}$ for the first two indices. 
This is why the stau exchange is forbidden in the $s$-channel diagram 
in $P\to \tau \nu_\tau$. 
%%% paragraph 

%%% paragraph 
We first study the contribution of $t$-channel squark exchange in 
$D_s \to \tau \nu_\tau$. 
When the sbottom exchange diagram is dominant, the contribution 
to the parameter $r_{D_s}$ is given by a coupling $\lambda'_{323}$, while 
the parameter $r_{B^+}$ is given by a product 
$\lambda'_{313}\lambda'^*_{333}$.  
In Fig.~\ref{tch_graph}, we show the sbottom contribution to 
$r_{D_s}$ for $D_s \to \tau \nu_\tau$ (left) and 
$r_{B^+}$ for $B^+ \to \tau \nu_\tau$ (right) as a function 
of $|\lambda'_{323}|^2$ and $\lambda'_{313}\lambda'^*_{333}$, 
respectively.  
The horizontal lines denote constraints on 
$r_{D_s}$ from (\ref{cleo2009}) and (\ref{fdqcd}), 
and $r_{B^+}$ from (\ref{expB}) and (\ref{utfit})
\begin{subequations}
\begin{eqnarray}
r_{D_s} &=& 1.07 \pm 0.04,  
\label{rparamexp_ds}
\\
r_{B^+} &=& 1.44 \pm 0.23. 
\label{rparamexp_bp}
\end{eqnarray}
\end{subequations}
%------------
From Fig.~\ref{tch_graph}, we find that the $t$-channel contribution 
constructively interferes with the $W$-boson exchange, i.e., 
$r_{D_s}, r_{B^+} \ge 1$. 
Taking account of eqs.(\ref{rparamexp_ds}) and (\ref{rparamexp_bp}), 
constraints on the RPV couplings at 1-$\sigma$ level are given as 
\begin{eqnarray}
0.02 \simlt &|\lambda'_{323}|^2& \simlt 0.07,  
\label{dsconst}
\\
0.0006 \simlt &\lambda'_{313}\lambda'^*_{333}& \simlt 0.0017. 
\label{bpconst}
\end{eqnarray}
The allowed RPV couplings for $B^+ \to \tau \nu_\tau$ 
(\ref{bpconst}) 
is smaller than that for $D_s \to \tau \nu_\tau$ 
(\ref{dsconst}) 
by few orders of magnitude. 
This is because the parameter $r_P$ (\ref{rparam}) 
accounts for the relative size of new physics contribution against 
for a CKM matrix element. 
Note that the CKM matrix element in $r_{B^+}$ is $V_{ub}\sim 10^{-3}$, 
while that in $r_{D_s}$ is $V_{cs} \sim 1$. 
Thus, the difference of magnitude between $V_{cs}$ and $V_{ub}$ explains 
the difference between (\ref{dsconst}) and (\ref{bpconst}).  
%%%

%%%
%--------
\begin{figure}
 \begin{center}
  \includegraphics[width=7cm]{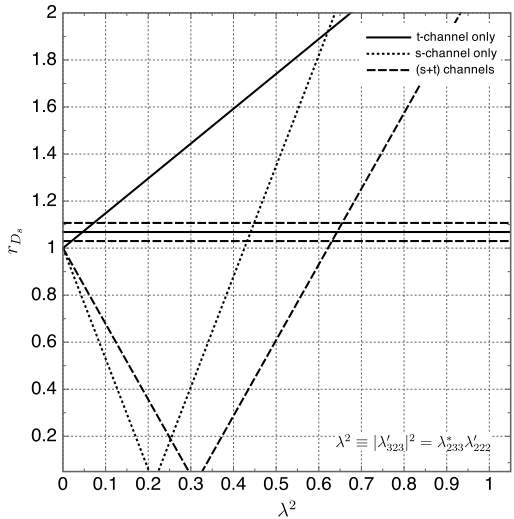}
  \caption{
Contribution of RPV interactions to the 
parameter $r_{D_s}$ as functions of 
the RPV couplings. 
Three curves correspond to the $t$-channel contribution (solid), 
the $s$-channel contribution (dotted) and the sum of $s$- and 
$t$-channel contributions (dashed). 
For comparison of contributions from each diagram, we fix 
the RPV couplings in both $t$- and $s$-channel diagrams to be 
equal, i.e., 
$\lambda^2 \equiv |\lambda'_{323}|^2 = \lambda'_{222}\lambda^*_{233}$. 
The horizontal lines denote 
the 1-$\sigma$ bound on $r_{D_s}$ from the experimental data. 
\label{dsstch} 
}
 \end{center}
\end{figure}
%--------

Next we study the $s$-channel slepton exchange. 
From (\ref{asp}), we find that the interference between the $s$-channel 
contribution and the $W$-boson exchange is destructive when the RPV 
couplings are real and positive. 
We show the $r_{D_s}$ parameter via the $s$-channel slepton exchange,
and the interference between the $s$- and $t$-channel exchanges 
in Fig.~\ref{dsstch}. 
Note that 
the $t$-channel contribution is proportional to $|\lambda'_{323}|^2$ 
while the $s$-channel contribution is  
$\lambda'_{222}\lambda_{233}^*$. 
The solid curve represents the $t$-channel contribution which is 
obtained by setting $\lambda^2 = |\lambda'_{323}|^2$ and   
$\lambda'_{222}\lambda^*_{233}=0$. 
On the other hand, the dotted curve denotes the $s$-channel contribution
which is obtained by setting the $t$-channel coupling to zero, i.e., 
$\lambda^2 = \lambda'_{222}\lambda^*_{233}$ 
and $|\lambda'_{323}|^2=0$. 
The sum of the $t$- and $s$-channel diagrams is given 
by the dashed curve, where we fix the RPV couplings in both $t$- and 
$s$-channel diagrams to be equal,  
$\lambda^2 = |\lambda'_{323}|^2=\lambda'_{222}\lambda^*_{233}$ 
for comparison of contributions from each diagram.  
It is clear that, when 
$\lambda'_{222}\lambda^*_{233}>0$, 
the $s$-channel slepton exchange diagram destructively interferes with 
both the SM $W$-boson and $t$-channel squark exchange diagrams. 
Therefore, the squark ($t$-channel) and slepton ($s$-channel)
contributions may be cancelled each other in some parameter space. 
In the dashed line which represents the sum of $s$- and $t$-channel
the $r_{D_s}$ parameter decreases from unity and becomes zero 
(i.e., $G_F V_{cs} + A_t^{D_s} \approx -A_s^{D_s}$) around 
$\lambda^2 \sim 0.3$.  
For $\lambda^2 \simgt 0.3$, the $s$-channel contribution eventually 
dominates over the $W$-boson and $t$-channel squark contribution  
($G_F V_{cs} + A_t^{D_s} \ll |A_s^{D_s}|$) and the $r_{D_s}$ parameter 
increases with $\lambda^2$, 
which satisfies the experimental constraint when $\lambda^2 \sim 0.65$. 
From Fig.~\ref{dsstch} we find the relation $A_t^{D_s} < A_s^{D_s}$
holds and this can be understood as follows.  
When the RPV couplings and sparticle masses are same in both 
the $t$-channel contribution $A_t^P$ (\ref{atp}) and 
the $s$-channel contribution $A_s^P$ (\ref{asp}), 
the relative magnitudes of two contributions are determined by 
their coefficients, $\frac{1}{4\sqrt{2}}$ for the $t$-channel 
and 
$\frac{1}{2\sqrt{2}}\frac{m_P}{m_{l_i}}
\frac{m_P}{m_{U_a}+ m_{D_b}}$ for the $s$-channel. 
Note that the ratio $\frac{m_P}{m_{U_a}+ m_{D_b}}$ is of order unity 
for $P=D_s$ or $B^+$. 
On the other hand, the $s$-channel contribution $A_s^P$ could be
enhanced by the ratio $\frac{m_P}{m_{l_i}}$. 
For $l_i=\tau$, the ratio is $\frac{m_P}{m_\tau}\sim 1$ 
for $P=D_s$ and $\sim 3$ for $P=B^+$. 
Therefore, the size of $A_s^P$ is about $2(8)$ times larger than 
$A_t^P$ for $D_s (B^+)$ when the RPV couplings and the sparticle 
masses are common. 
It should be mentioned that the $s$-channel contribution 
$A_s^{D_s}$ is considerably larger for $l=\mu(e)$ 
than for $l=\tau$ due to small lepton mass. 
%--------
\begin{figure}
 \begin{center}
  \includegraphics[width=7cm]{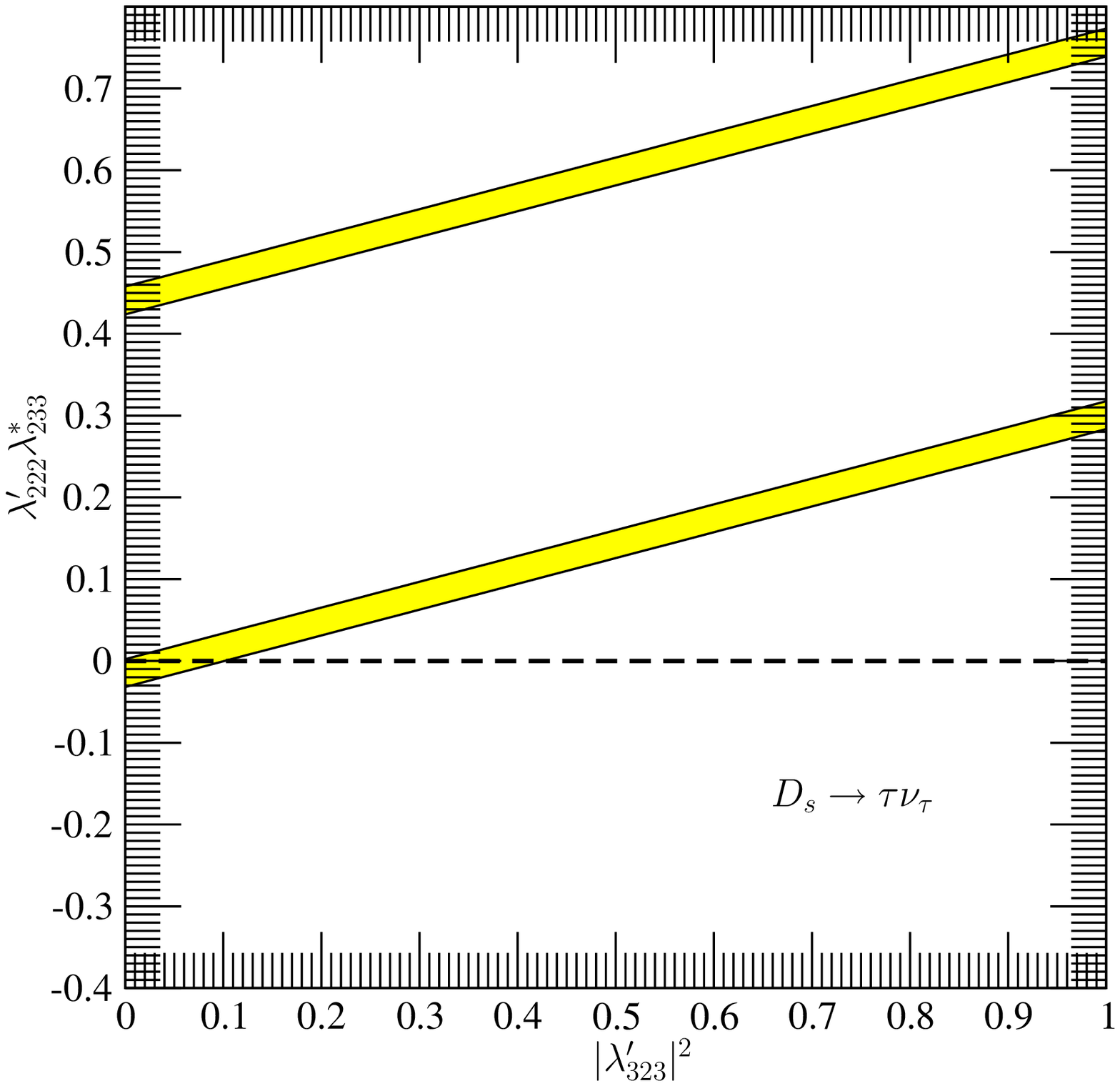}
  \includegraphics[width=7cm]{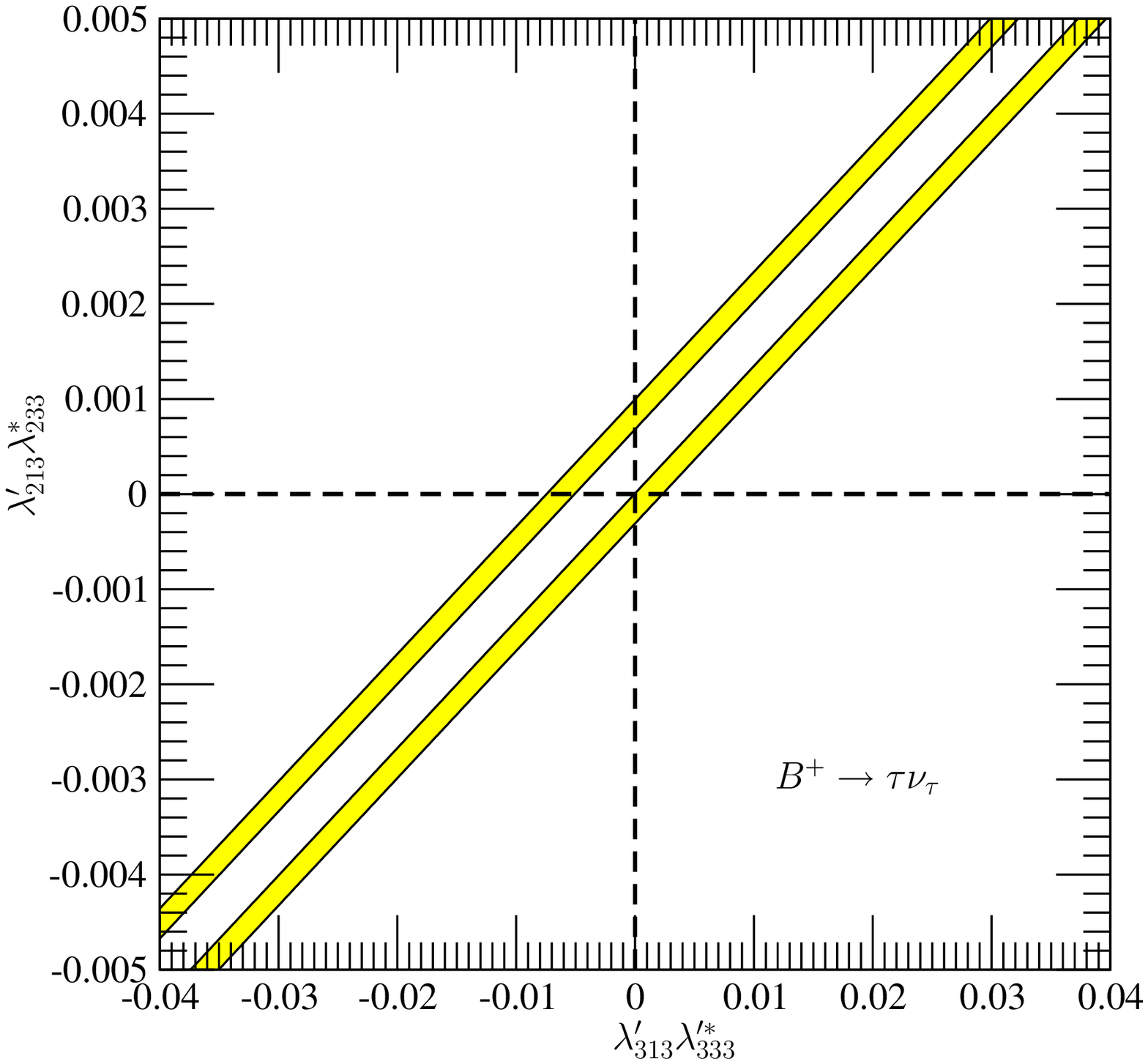}
  \caption{
Constraints on the RPV couplings from $D_s \to \tau \nu_\tau$ (left) 
and $B^+ \to \tau \nu_\tau$ (right). 
The horizontal axis represents the RPV couplings for $t$-channel while 
the vertical axis denotes the couplings for $s$-channel. 
The bands with solid line correspond to the 2-$\sigma$ allowed range 
for the $r_{D_s}$ (left) and $r_{B^+}$ (right) parameters, respectively.  
In each figure, the inner lines correspond to the 2-$\sigma$ lower 
bounds while the outers are the 2-$\sigma$ upper bounds on $r_{D_s}$ 
and $r_{B^+}$, respectively. 
\label{diffcouplings}}
\end{center}
\end{figure}
%--------

%--------
In Fig.~\ref{diffcouplings}, we show constraints on the RPV couplings 
from $D_s \to \tau \nu_\tau$ (left) and $B^+ \to \tau \nu_\tau$ (right).  
The horizontal axis represents the RPV couplings for $t$-channel diagram 
while the vertical axis denotes the couplings for $s$-channel diagram.  
The bands with solid lines correspond to the 2-$\sigma$ allowed range 
for $r_{D_s}$ (\ref{rparamexp_ds}) and $r_{B^+}$ (\ref{rparamexp_bp}). 
In each figure, the inner solid lines correspond to the 2-$\sigma$ 
lower bounds while the outers are the 2-$\sigma$ upper bounds on
$r_{D_s}$ and $r_{B^+}$, respectively. 
From Fig.~\ref{diffcouplings}, we find that the $s$-channel couplings 
have positive  
correlations with the $t$-channel couplings. 
This is because the interference between the $s$- and $t$-channel 
contributions is destructive. 
For $D_s \to \tau \nu_\tau$, since the $t$-channel coupling is always 
positive ($|\lambda'_{323}|^2 \ge 0$), not only the magnitude but also 
the sign of $s$-channel coupling $\lambda'_{222}\lambda^*_{233}$ 
is strongly constrained. 
For negative $\lambda'_{222}\lambda^*_{233}$, 
the $t$-channel coupling $|\lambda'_{323}|^2$ should be smaller than 
$0.12$ and, then, $-0.04 \simlt \lambda'_{222}\lambda^*_{233} \le 0$ 
is experimentally allowed in the 2-$\sigma$ level. 
For $B^+ \to \tau \nu_\tau$, 
the $s$- and $t$-channel couplings with opposite 
signs are strongly constrained. 
As can be seen in Fig.~\ref{diffcouplings}, when the $t$-channel 
coupling is positive  
($\lambda'_{313}\lambda'^*_{333} \ge 0$), the negative $s$-channel
coupling is constrained to be  
$-0.0004 \simlt \lambda'_{213}\lambda^*_{233}\le 0$. 
Although the leptonic decays of $D_s$ and $B^+$ mesons are useful 
to constrain the sign of the relevant RPV couplings, 
the size of the couplings cannot be restricted because of the 
cancellation among the diagrams. 
However, the correlations among the RPV couplings as shown in 
Fig.~\ref{diffcouplings} may be a good information to test the
$R$-parity violating SUSY-SM at the direct search experiments such as
LHC, because some RPV couplings could be large simultaneously and it 
may lead to observation of several productions or decay processes 
due to the RPV interactions. 
%--------
\begin{figure}
 \begin{center}
  \includegraphics[width=7cm]{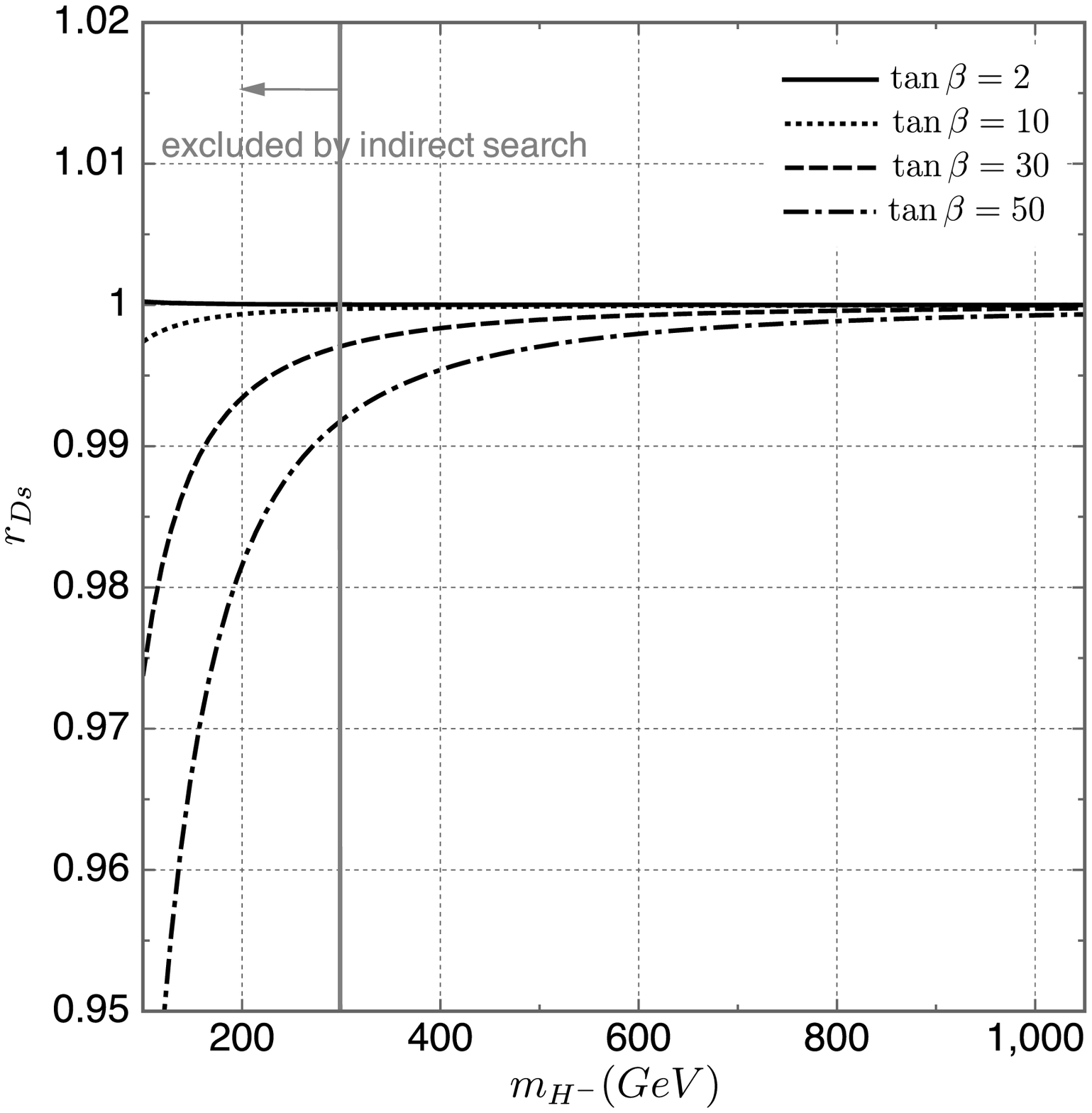}
  \includegraphics[width=7cm]{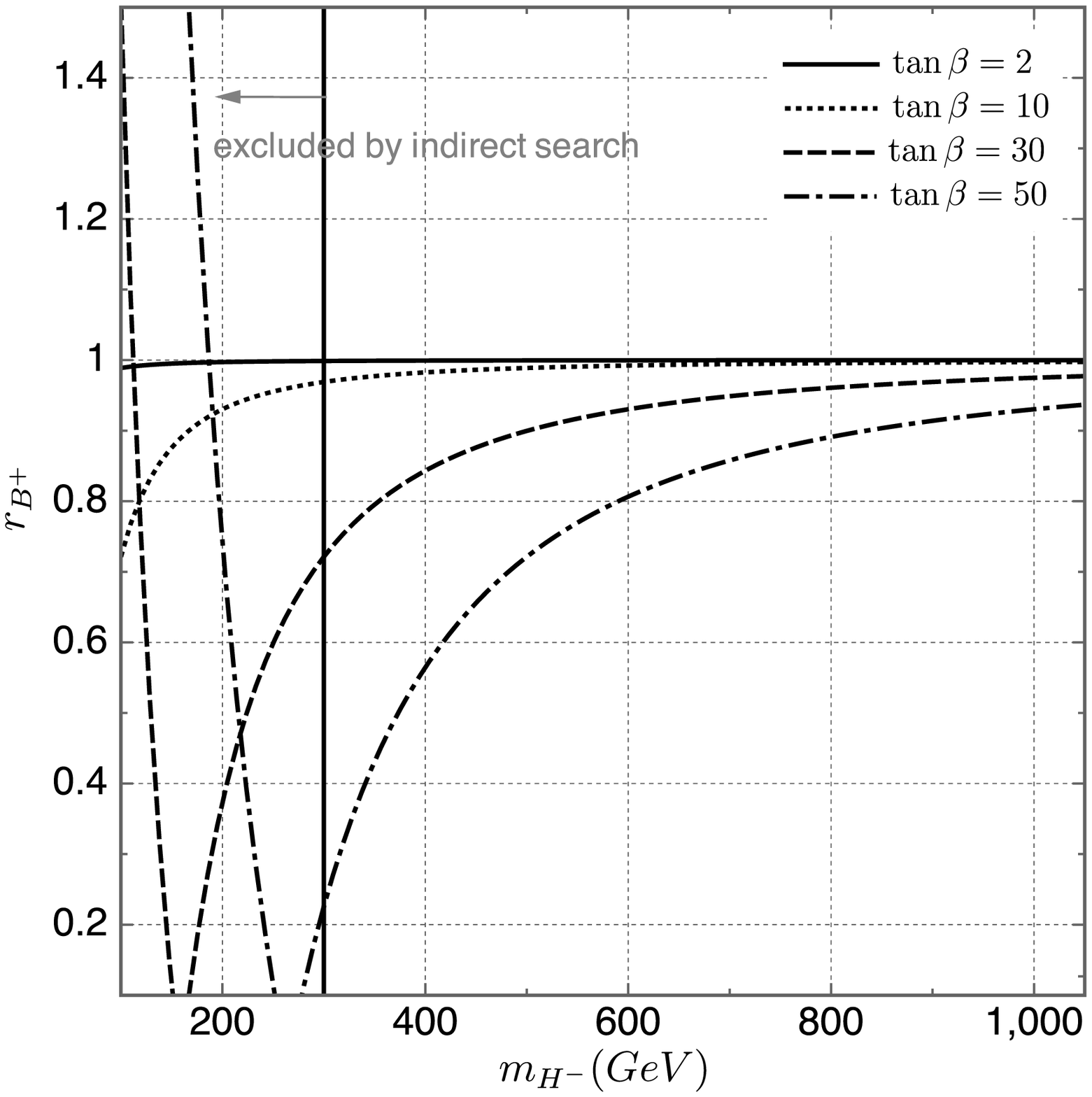}
  \caption{
The charged Higgs contributions to 
$r_{D_s}$ (left) and $r_{B^+}$ (right) as functions of 
the charged Higgs boson mass $m_{H^-}$. 
The four curves (solid, dotted, dashed and dot-dashed) 
correspond to $\tan\beta=2,10,30$ and $50$, respectively. 
The vertical line denotes the lower bound on $m_{H^-}>295{\rm GeV}$ 
from the $b\to s\gamma$ decay~\cite{Misiak:2006zs}. 
\label{charged}
}
 \end{center}
\end{figure}
%--------

\begin{figure}
 \begin{center}
  \includegraphics[width=7cm]{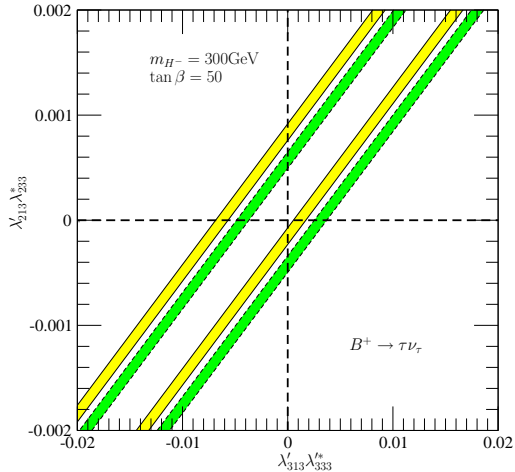}
  \caption{
Constraints on the RPV couplings from $B^+ \to \tau \nu_\tau$. 
The bands with solid and dotted lines correspond to the 1-$\sigma$ 
allowed range for the $r_{B^+}$ parameter with and without the 
contribution from charged Higgs exchange, respectively. 
The contribution of charged Higgs boson is estimated for 
$m_{H^-}=300~{\rm GeV}$ and $\tan\beta=50$. 
\label{cont_higgs}
}
 \end{center}
\end{figure}

%--------
In the MSSM with RPV couplings, in addition to the contributions via 
squark and slepton exchanges, the charged Higgs boson $H^-$ also affect 
the leptonic decay of a pseudo scalar meson through the $s$-channel diagram.   
We show the charged Higgs contributions to $r_{D_s}$ and $r_{B^+}$ 
as functions of the mass $m_{H^-}$ in Fig.~\ref{charged}. 
In each figure, four curves are obtained for $\tan\beta=2,10,30$ and 
50. 
The vertical line denotes the lower bound on the mass of charged Higgs 
boson from the experimental data of $b \to s \gamma$, 
$m_{H^-}> 295~{\rm GeV}$~\cite{Misiak:2006zs}\footnote{
This constraint has been obtained on Type-II two Higgs 
doublet model (THDM-II). Although the Higgs sector in the MSSM 
has the same structure with THDM-II, the contributions of 
$H^-$ to $b \to s\gamma$ could be canceled with those from 
charginos. Therefore, the constraint on $m_{H^-}$ which we adopted here 
corresponds to the decoupling limit of chargino so that 
it may be conservative. }. 
It is easy to see that the charged Higgs contribution to 
$D_s \to \tau \nu_\tau$ is marginal  
(smaller than 1\%) for $m_{H^-} > 295~{\rm GeV}$. 
On the other hand, the contribution to $B^+ \to \tau \nu_\tau$ could 
be as large as $80\%$ for $\tan\beta=50$. 
However, it destructively interferes with the $W$-boson exchange so that
the charged Higgs contribution is disfavored from the current 
experimental data of $B^+ \to \tau \nu_\tau$. 
Thus, the charged Higgs contribution cannot explain the deviation 
between the data and the SM prediction in the leptonic decay 
of $B^+$ meson.  
Comparison of constraints on the RPV couplings from 
$B^+ \to \tau \nu_\tau$ with and without the charged Higgs exchange 
is shown in Fig.~\ref{cont_higgs}. 
The bands with solid and dotted lines correspond to the 1-$\sigma$ 
allowed range for $r_{B^+}$ with and without the charged 
Higgs exchange, respectively. 
The contribution of charged Higgs exchange is estimated for 
$m_{H^-}=300~{\rm GeV}$ and $\tan\beta=50$, which is a parameter set 
to give most sizable contribution to  
$B^+ \to \tau \nu_\tau$ under the experimental constraints from 
$b \to s \gamma$ as shown in Fig.~\ref{charged}. 
The contribution of charged Higgs boson slightly alters the allowed 
region of the RPV couplings. 
For example, when the $s$-channel RPV couplings are zero 
($\lambda'_{213}\lambda^*_{233}=0$), the allowed range of RPV couplings 
with the charged Higgs contribution shifts about factor two or three 
from that without charged Higgs boson. 
%--------
%--------
\section{Summary}
We have investigated the leptonic decays of $D_s$ and $B^+$
mesons in $R$-parity violating supersymmetric SM. 
The experimental data of leptonic decays 
$D_s \to \tau \nu_\tau$ and $B^+ \to \tau \nu_\tau$ show 
about 2.4 and 2.5-$\sigma$ deviations from the SM (Lattice QCD) 
predictions.  
We found the parameter space of the R-parity violating supersymmetric 
SM to explain the above deviations. 
It was shown that the interference between the $s$-channel slepton 
exchange and $t$-channel squark exchange diagrams could be 
either constructive or destructive, owing to a choice of relative 
sign between the RPV couplings in two diagrams. 
We also found that when the relative sign of the RPV couplings between  
the $s$- and $t$-channels is opposite, the allowed parameter region 
is strongly restricted from the experimental data. 
For example, in case of $D_s \to \tau \nu_\tau$, 
the negative $s$-channel coupling $\lambda'_{222}\lambda^*_{233} \le 0$  
is allowed only when the $t$-channel coupling is 
$|\lambda'_{323}|^2 \simlt 0.012$. 
In case of $B^+ \to \tau \nu_\tau$, the RPV couplings 
$\lambda'_{213}\lambda^*_{233} < 0$ 
in $s$-channel and 
$\lambda'_{313}\lambda'^*_{333}> 0$ in $t$-channel 
are constrained to be less than $10^{-3}$. 
The charged Higgs contribution always destructively interferes with 
the SM $W$-boson contribution. 
Taking account of constraints on the charged Higgs mass from the 
$b \to s\gamma$ decay, we found that 
the charged Higgs contribution to the parameter $r_{D_s}$ is marginal 
while that to $r_{B^+}$ could be sizable for large $\tan\beta$ because
of the enhancement of the bottom-Yukawa coupling. 
We presented how the constraints on RPV couplings are altered with 
and without charged Higgs contribution in $B^+ \to \tau \nu_\tau$. 

A distinct feature of our work from the previous studies is that the 
RPV couplings related to $D_s \to \tau \nu_\tau$ and 
$B^+ \to \tau \nu_\tau$ could be sizable {\it simultaneously} due to 
the positive correlation between the $s$- and $t$-channel diagrams. 
Since the expected sensitivity of the RPV couplings at LHC is, e.g.,  
$0.1 -0.01$ for $\lambda'_{ijk}$ from the single sparticle production 
events with the integrated luminosity 
$\int dt {\cal L}=30{\rm fb}^{-1}$~\cite{Barbier:2004ez}, 
the allowed parameter space of the RPV 
couplings which is found in our study will be covered and 
our scenario to explain the deviation in the leptonic decays of $D_s$ and 
$B^+$ mesons using the RPV couplings could be tested.

%--------------------------------------------

\end{document}